\documentclass[ aps,%
12pt,%
floatfix,%
final,%
notitlepage,%
oneside,%
onecolumn,%
nobibnotes,%
nofootinbib,%
superscriptaddress,%
showpacs,%
]%
{revtex4}

\usepackage{graphicx}
\usepackage{amssymb}
\usepackage{hyperref}

\newcommand{\MM}{\mathcal{M}}
\newcommand{\A}{\mathcal{A}}
\newcommand{\B}{\mathcal{B}}
\newcommand{\Q}{\mathcal{Q}}

\newcommand{\Br}{\mathrm{Br}}

\newcommand{\SPS}{\text{SPS}}
\newcommand{\DPS}{\text{DPS}}

\newcommand{\GeV}{\text{GeV}}
\newcommand{\nb}{\text{nb}} \newcommand{\pb}{\text{pb}}
\newcommand{\mub}{\mu\text{b}} \newcommand{\mb}{\text{mb}}
\newcommand{\fb}{\text{fb}}
\newcommand{\CO}{\text{CO}}\newcommand{\NLO}{\text{NLO}}

\begin{document}

\title{Simultaneous  production of charmonium and bottomonium mesons at LHC}

\author{A.~K.~Likhoded}
\email{Anatolii.Likhoded@ihep.ru}
\affiliation{Institute for High Energy Physics, Prtovino, Russia}
\affiliation{Moscow Institute of Physics and Technology, Dolgoprudny, Russia}

\author{A.~V.~Luchinsky}
\email{alexey.luchinsky@ihep.ru}
\affiliation{Institute for High Energy Physics, Prtovino, Russia}
\affiliation{Institute of Theoretical and Experimental Physics, Moscow, Russia}

\author{S.~V.~Poslavsky}
\email{stvlpos@mail.ru}
\affiliation{Institute for High Energy Physics, Prtovino, Russia}
\affiliation{Institute of Theoretical and Experimental Physics, Moscow, Russia}

\begin{abstract}
Inclusive production of $\Upsilon J/\psi$ pair in proton-proton interation at LHCb is considered. This process is forbidden at leading order of perturbation theory, so such channels as double parton scattering, $\chi_b\chi_c$ pair production with subsequent radiative decays of $P$-wave quarkonia, contributions of color-octet states, and NLO corrections are studied in details. For all these channels we present theoretical predictions of total cross sections at LHCb and distributions over different kinematical variables. According to presented in the paper results, double parton interaction gives main contribution to the cross section of the considered reaction.
\end{abstract}

\maketitle

\section{Introduction}

Inclusive double production of heavy quarkonia was observed in NA3 experiment, where production of two $J/\psi$ mesons in $\pi N$ interaction was studied. The cross section of double $J/\psi$ production was suppressed by about three orders of magnitude in comparison with single $J/\psi$. Such suppression can be easily explained at leading order (LO) of perturbative QCD \cite{Kartvelishvili:1984ur,Humpert:1983yj}. In low energy hadronic collisions quark-antiquark subprocess  $q\overline{q}\to J/\psi J/\psi$ gives main contribution to the considered reaction, while in the case of high energy interactions gluonic subprocesses $gg\to J/\psi J/\psi$ are dominant. Leading order $O(\alpha_s^4)$ QCD predictions and subsequent next to leading order (NLO) corrections \cite{Lansberg:2013qka} are in  good agreement with experimental results, obtained at LHC \cite{Aaij:2014rms, Khachatryan:2014iia, CMS:2013pph, Aaij:2011yc, Berezhnoy:2012xq, Berezhnoy:2011xy}. It is clear, on the other hand, that at LHC energies the 
gluoinc luminosity is large, so production of two $J/\psi$ mesons in two independent partonic interactions (so called double parton scattering, DPS) is possible. In papers \cite{Baranov:2011ch, Novoselov:2011ff} it is shown, that cross sections calculated in DPS approximation are comparable with the ones obtained in single parton scattering (SPS) approximation. The drawback of DPS is poor knowledge of double parton distribution functions. If one accepts the simplest model and assumes, that this function is just the product of two single-parton distribution functions, the following simple relation holds:
\begin{eqnarray*}
\text{\ensuremath{\sigma_{\DPS}^{AB}}} & \approx & \frac{\sigma_{\SPS}^{A}\sigma_{\SPS}^{B}}{\sigma_{eff}}.
\end{eqnarray*}
Using values of single particle production cross sections $\sigma_{SPS}^{A,B}$ and effective cross section $\sigma_{eff}\approx14.5\ \nb$ from CDF and D0 data \cite{Abe:1997xk,Abazov:2009gc,Snigirev:2003cq}, it is easy to calculate the cross section of double $J/\psi$ production via DPS:
\begin{eqnarray*}
\sigma_{\DPS}^{J/\psi J/\psi} & \approx & 2\ \nb.
\end{eqnarray*}
This value can be compared with SPS cross section, that is calculated at LO $O(\alpha_s^4)$ perturbative QCD:
\begin{eqnarray*}
\sigma_{\SPS}^{J\psi J\psi} &\approx&4\ \nb.
\end{eqnarray*}

As for double $\Upsilon(1S)$ meson production, presented in paper \cite{Berezhnoy:2012tu} estimates show, that cross section of this process is about three orders of magnitude smaller than cross section of double $J/\psi$ production. For energy $\sqrt{s}=7$ TeV at LHCb we have
\begin{eqnarray*}
\sigma_{\SPS}^{\Upsilon\Upsilon} & \approx & 8.7\ \pb,\quad
\sigma_{\DPS}^{\Upsilon\Upsilon}  \approx  0.4\ \pb.
\end{eqnarray*}
It can be clearly seen, that while for double $J/\psi$ production DPS and SPS cross sections are comparable, in the case of $\Upsilon(1S)\Upsilon(1S)$ final state DPS cross section is suppressed by more than order of magnitude.

Question concerning situation in $\Upsilon J/\psi$ production arises naturally. At leading $O(\alpha_s^4)$ order in color-singlet approximation this process is forbidden, so the following relation holds:
\begin{eqnarray}
\sigma_{\SPS}^{\Upsilon J/\psi} & \ll & \sigma_{\SPS}^{\Upsilon\Upsilon}\ll\sigma_{\SPS}^{J/\psi J/\psi}.\label{eq:sigmaRel}
\end{eqnarray}
The reason is that at leading $O(\alpha_s^4)$ order for $J/\psi J/\psi$ and $\Upsilon\Upsilon$ processes diagrams with $t$-channel quark exchange are allowed, that saturate low invariant mass region and lead to rapid decrease of the cross section with the increase of this invariant mass, $\hat\sigma\sim1/\hat{s}^{2}$. At the same order of magnitude the contributions to $\Upsilon J/\psi$ production are absent, but production via radiative decays of $\chi_b\chi_c$ state is allowed. So, it is interesting to study the modification of relation (\ref{eq:sigmaRel}) with these processes taken into account.

In the next section cross sections of $\chi_b\chi_c$ production are considered. Section III is devoted to calculation of NLO $O(\alpha_s^6)$ corrections and color-octet contributions to the cross section of the process $gg\to\Upsilon J/\psi$. In section IV we estimate cross section of inclusive $\Upsilon J/\psi$ production at LHCb and compare them with DPS contribution.

\section{$gg\to\chi_{c}\chi_{b}$}

In the framework of partonic model the cross section of inclusive double heavy quarkonia production $\mathcal{Q}_{1}\mathcal{Q}_{2}$ is written as a convolution of partonic distribution functions (at LHC energies gloun-gluon interaction gives main contribution) and cross section of the hard subprocess. The latter can be calculated using NRQCD formalism \cite{Bodwin:1994jh}, when the cross section is written as a series over a relative quark velocity (in the following we will restrict ourself to leading order). In such approach the fock column of the heavy quarkonium contains both colour singlet (CS) and colour octet (CO) components. According to NRQCD scaling rules contributions of CS and CO mechanisms should be comparable, but according to presented in \cite{Likhoded:2012hw,Likhoded:2013aya,Likhoded:2014gpa,Likhoded:2014kfa} analysis, where experimental results on single $\chi_c$ meson production at LHC were considered, CS contributions dominate. For this reason in the following we will neglect CO 
components of $P$-wave heavy quarkonia.

\begin{figure}
\includegraphics{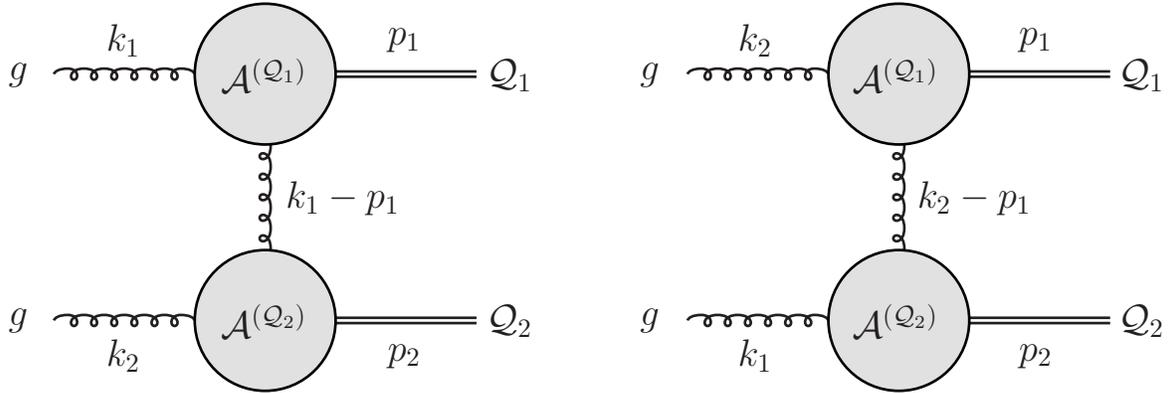}
\caption{Feynman diagrams of $gg\to\chi_{c}\chi_{b} subprocess$\label{diags2chi}}
\end{figure}

At the lading order of perturbation theory hard cross sections of the considered processes are described by shown in Fig.\ref{diags2chi} diagrams and the corresponding amplitude can be written in the form
\begin{eqnarray}
\MM\left(gg\to\Q_{1}\Q_{2}\right) & = & \epsilon_{1}^{\alpha}\epsilon_{2}^{\beta}\left\{ \frac{1}{\hat{t}}\A_{\alpha\mu}^{\left(\Q_{1}\right)}\left(k_{1},p_{1}-k_{1}\right)\A_{\beta\mu}^{\left(\Q_{2}\right)}\left(k_{2},p_{2}-k_{2}\right)+\right.\nonumber \\
 &  & \left.\frac{1}{\hat{u}}\A_{\beta\mu}^{\left(\Q_{1}\right)}\left(k_{2},p_{1}-k_{2}\right)\A_{\alpha\mu}^{(\Q_{2})}\left(k_{1},p_{2}-k_{1}\right)\right\} ,\label{eq:amp}
\end{eqnarray}
where $k_{1,2}$, $p_{1,2}$ are momenta of initial gluons and final quarkonia respectively, $\hat{s}=(k_{1}+k_{2})^{2}$, $\hat{t}=(p_{1}-k_{1})^{2}$, $\hat{u}=(k_1-p_2)^2$ are Mandelstam variables of the partonic subprocess,  $\epsilon_{1,2}$ are polarisation vectors of initial gluons, and $\A_{\alpha\beta}^{(\Q)}(k,q)$ is the effective vertex that describe quarkonium $\Q$ production in fusion of two (possibly virtual) gluons with momenta $k$, $q$. Using NRQCD formalism it is easy to obtain the expressions for these vertices. In the case of scalar and axial quarkonia, for example, we have
\begin{eqnarray}
\A_{\alpha\beta}^{(\chi_{0})}(k,q) & = & -\sqrt{\frac{2\left\langle O_1^{\chi_Q}\right\rangle }{3M^{3}}}\frac{g_{s}^{2}\delta_{ab}}{\left(M^{2}-k^{2}-q^{2}\right)^{2}}\nonumber \\
 &  & \left\{ -4k^{2}q^{\alpha}q^{\beta}-4q^{2}k^{\alpha}k^{\beta}+2\left(3M^{2}-k^{2}-q^{2}\right)\left(k^{\alpha}q^{\beta}+k^{\beta}q^{\alpha}\right)-\right.\nonumber \\
 &  & g^{\alpha\beta}\left((k^{2}-q^{2})^{2}+M^{2}\left(3M^{2}-4k^{2}-4q^{2}\right)\right),\label{eq:A0}\\
\A_{\alpha\beta}^{(\chi_{1})}(k,q) & = & \frac{4\sqrt{\left\langle O_1^{\chi_Q}\right\rangle }}{3M^{5/2}}\frac{g_{s}^{2}\delta_{ab}}{\left(M^{2}-k^{2}-q^{2}\right)}\nonumber \\
 &  & \left\{ \left(k^{2}-q^{2}\right)\left(M^{2}-k^{2}-q^{2}\right)\left(e^{\alpha\beta\mu\nu}k_{\nu}+e^{\alpha\beta\mu\nu}q_{\nu}\right)-\right.\nonumber \\
 &  & 2\left[\left(M^{2}-k^{2}-q^{2}\right)k^{\alpha}-\left(k^{2}+q^{2}\right)q^{\alpha}\right]e^{\beta\mu\nu\gamma}k_{\nu}q_{\gamma}+\nonumber \\
 &  & \left.2\left[\left(M-k^{2}-q^{2}\right)q^{\beta}-\left(k^{2}+q^{2}\right)k^{\beta}\right]e^{\alpha\mu\nu\gamma}k_{\nu}q_{\gamma}\right\} ,\label{eq:A1}
\end{eqnarray}
where color-singlet NRQCD matrix element
\begin{eqnarray}
\left\langle O_1^{\chi_{Q}}\right\rangle  & = & \frac{3N_{c}}{2\pi}\left|R'_{QQ}(0)\right|^{2}\label{eq:Ochi}
\end{eqnarray}
was introduced. Its numerical value can be obtained, for example, with the help of potential models or from the analysis of available experimental data. In the case of tensor quarkonium the relation similar to (\ref{eq:A0}), (\ref{eq:A1}) was also obtained, but it is too complex to present it here. It should be noted, however, that in the case of axial meson the effective vertex tends to zero in the limit $k^{2}=q^{2}=0$. The reason for such behaviour is that, in agreement with Landau-Yang theorem, interaction of axial meson with two massless gluon is forbidden.

From presented above amplitude it is easy to get the cross section of the partonic reaction: 
\begin{eqnarray*}
\frac{d\hat{\sigma}}{d\hat{t}} & = & \frac{1}{128\pi\hat{s}^{2}}\left|\MM\right|^{2}.
\end{eqnarray*}
These calculations were performed in REDBERRY computer algebra system \cite{redberry} and the resulting expressions are, unfortunately, too large to present them here (in computer readable form they can be found in supplementary files for this arXiv preprint). One can, however, make a simple estimate in high energy region $\sqrt{\hat{s}}\gg M_{J/\psi}$ $M_{\Upsilon}$. From relation (\ref{eq:amp}) it is clear that configurations with small transverse momentum $p_{T}\approx\sqrt{\hat{t}\hat{u}/\hat{s}}$ give main contributions in this region, so, for example, for small $\hat t$ one can take only first term in the amplitude (\ref{eq:amp}). The squared amplitude in this approximation can be written in the form
\begin{eqnarray}
\left|\MM\right|^{2} & \approx & \frac{1}{\hat{t}^{2}}\B_{\mu\nu}^{(\chi_{cJc})}(k_{1},p_{1}-k_{1})\B_{\mu\nu}^{(\chi_{bJ_{b}})}(k_{2},p_{2}-k_{2}),\label{eq:MB}
\end{eqnarray}
where notation
\begin{eqnarray}
\B_{\mu\nu}^{(\Q)}(k,q) & = & \sum_\mathrm{pol}\A_{\alpha\mu}^{(\Q)}(k,q)\A_{\alpha\nu}^{(\Q)*}(k,q)\label{eq:B}
\end{eqnarray}
is used. If we use the leading term over $1/\hat t$ and set $\hat{t}\approx0$ in expressions (\ref{eq:B}) (in high energy region this is allowed kinematically), in the case of tensor and axial quarkonia we have
\begin{eqnarray}
\B_{\mu\nu}^{(\chi_{Q0})}(k,q) & =\frac{3}{4}\B_{\mu\nu}^{(\chi_{Q2})}(k,q)= & \frac{64\left\langle O_1^{\chi_Q}\right\rangle g_{s}^{4}}{3M^{3}}\left(g^{\mu\nu}-\frac{k^{\mu}q^{\nu}+k^{\nu}q^{\mu}}{(kq)}\right),
\label{eq:B02}
\end{eqnarray}
while for axial meson $\B^{(\chi_{1})}\approx0$. Using relations (\ref{eq:B02}) it is easy to obtain well known formulas for scalar and tensor quarkonia decay widths into two massless gluons \cite{Olsson:1984qe}:
\begin{eqnarray}
\Gamma\left(\chi_{Q0}\to2g\right) & = & 96\alpha_{s}^{2}\frac{\left|R'_{QQ}(0)\right|^{2}}{M^{4}},\label{eq:G0}\\
\Gamma\left(\chi_{Q2}\to2g\right) & = & \frac{128}{5}\alpha_{s}^{2}\frac{\left|R_{QQ}'(0)\right|^{2}}{M^{4}}.\label{eq:G2}
\end{eqnarray}
On the basis of presented above considerations one can obtain estimates for cross sections of scalar and tensor quarkonia production in high energy limit (see also \cite{Kiselev:1988mc}):
\begin{eqnarray*}
\hat{\sigma}\left(gg\to\chi_{bJ_{b}}\chi_{cJ_{c}}\right) & \sim & \left(2J_{c}+1\right)\left(2J_{b}+1\right)\frac{\Gamma\left(\chi_{cJ_{c}}\to2g\right)}{M_{\chi_{c}}}\frac{\Gamma\left(\chi_{bJ_{b}}\to2g\right)}{M_{\chi_{b}}}.
\end{eqnarray*}
The ratios of these cross sections are equal to
\begin{eqnarray}
\hat{\sigma}\left(\chi_{c2}\chi_{b2}\right):\hat{\sigma}\left(\chi_{c0}\chi_{b2}\right):\hat{\sigma}\left(\chi_{c2}\chi_{b0}\right):\hat{\sigma}\left(\chi_{c0}\chi_{b0}\right) & = & \frac{16}{9}:\frac{4}{3}:\frac{4}{3}:1.\label{eq:ratios}
\end{eqnarray}
If there is at least one axial meson in the final state, the corresponding cross sections should be suppressed.

\begin{figure}
\includegraphics[width=\textwidth]{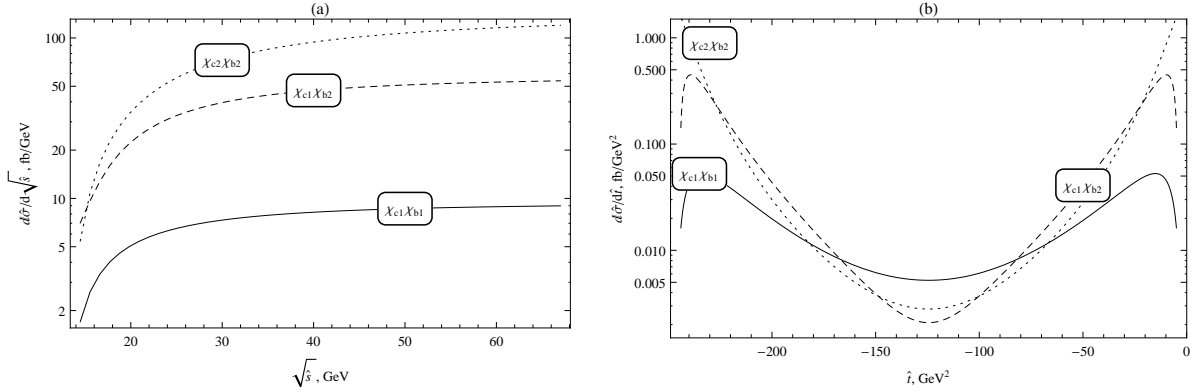}
\caption{
Dependence of partonic cross sections on $\sqrt{\hat s}$ (left panel) and $\hat t$ dependence at $'hat s=\hat s_0$  (right panel) for  $gg\to\chi_{b}\chi_{c}$ reactions. On both figures solid, dashed and dotted lines correspond to $gg\to\chi_{c1}\chi_{b1}$, $gg\to\chi_{c1}\chi_{b2}$, and  $gg\to\chi_{c2}\chi_{b2}$ subprocesses respectively.
\label{fig:hSigma}}
\end{figure}

For numerical calculations we should determine the value of NRQCD matrix element $\left\langle O_1^{\chi}\right\rangle $. At leading order over relative quark velocity one can use, for example, the derivative of quarkonia wave function at origin (see eq.(\ref{eq:Ochi})), that can be determined from solution of potential model. Alternatively  one can use relations (\ref{eq:G0}), (\ref{eq:G2}) to extract this parameter from experimentally known widths of scalar and tensor $\chi_Q$ mesons. For charmonia both methods give the same value $\left|R'_{cc}(0)\right|^{2}\approx0.075\ \GeV^{5}$. Presented in papers \cite{Likhoded:2013aya,Likhoded:2014gpa,Likhoded:2014kfa} analysis, however, shows, that experimental data are better described with a larger value
\begin{eqnarray*}
\left|R_{cc}'(0)\right|^{2} & = & 0.35\pm0.05\ \GeV^{5}.
\end{eqnarray*}
For bottomonium mesons the same approach gives \cite{Likhoded:2012hw}
\begin{eqnarray*}
\left|R'_{bb}(0)\right|^{2} & \approx & 1.0\ \GeV^{5}.
\end{eqnarray*}
These numbers will be used in the following. We will also neglect the fine splitting of heavy quarkonia masses, the following values will be used:
\begin{eqnarray*}
M_{\chi_{c}} & = & 3.5\ \GeV,\qquad M_{\chi_{b}}=9.9\ \GeV.
\end{eqnarray*}
The scale of the strong coupling constant $\alpha_{s}(\mu^{2})$ is chosen to be
\begin{eqnarray*}
\mu^{2} & = & \frac{M_{\chi_{c}}^{2}+M_{\chi_{b}}^{2}}{2}.
\end{eqnarray*}
Using presented above values it is easy to obtain predictions for partonic cross sections of $\chi$ meson production in gluon-gluon interaction. Integrated over total phase space partonic cross sections of the considered reactions at energy
\begin{eqnarray}
\hat{s} & = & \hat{s}_{0}=2(M_{\chi_{c}}+M_{\chi_{b}})^{2},\label{eq:s0}
\end{eqnarray}
for example, are listed in table \ref{tab:hSigma}. It is clearly seen that for ratios of scalar and tensor mesons production cross sections relations (\ref{eq:ratios}) are satisfied with pretty good accuracy. Dependence of some of these cross sections on $\sqrt{\hat s}$ and their distributions over $\hat{t}$ are shown in Fig.\ref{fig:hSigma}. One can see, that in the case of tensor meson production in low $\hat{t}$, $\hat{u}$ regions an increase caused by $\hat{t}$-channel gluon is observed. For axial meson production, on the other hand, in agreement with Landau-Yang theorem such increase is absent.

\begin{table}
\bigskip
\begin{tabular}{|c|c|c|c|}
\hline 
$\Q_{1}/\Q_{2}$ & $\chi_{c0}$ & $\chi_{c1}$ & $\chi_{c2}$\tabularnewline
\hline 
\hline 
$\chi_{b0}$ & 16.3 & 14.8 & 21.4\tabularnewline
\hline 
$\chi_{b1}$ & 2.1 & 4.6 & 3.8\tabularnewline
\hline 
$\chi_{b2}$ & 21.4 & 19.6 & 29.2\tabularnewline
\hline 
\end{tabular}
\caption{Total cross sections of the partonic reactions $gg\to\chi_{cJ_{c}}\chi_{bJ_{b}}$ (in fb) at $\hat{s}=\hat{s}_{0}$\label{tab:hSigma}}
\end{table}

\begin{figure}
\includegraphics[width=1\textwidth]{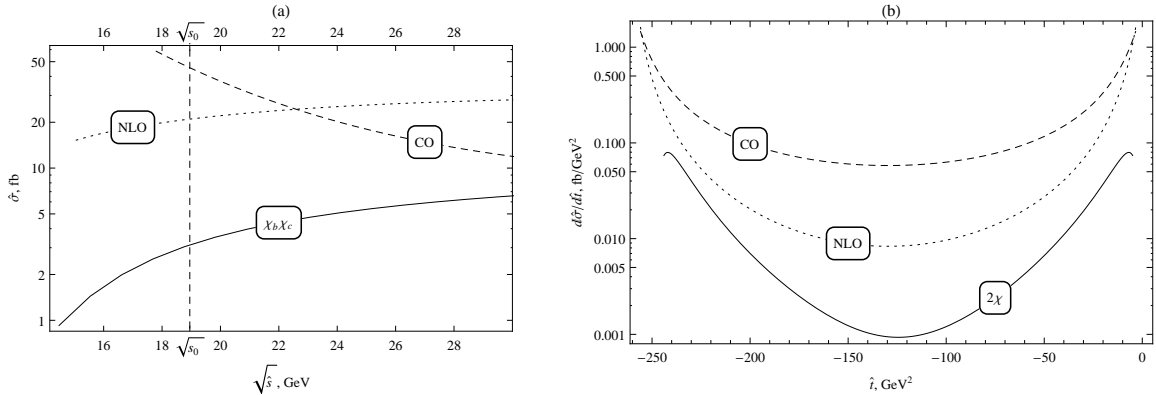}
\caption{
Dependence of cross section of $\Upsilon J/\psi$ production in partonic subprocess (left panel) and its $\hat t$-distribution at $\hat s=\hat{s}_0$ (right panel). Solid, dotted, and dashed lines correspond to $\chi_b\chi_c$, NLO, and color-octet channels respectively.\label{fig:hplotNLO}}
\end{figure}

After radiative decays of $\chi_{c,b}$ mesons into vector quarkonia reactions $gg\to\chi_{b}\chi_{c}$ give contributions into cross sections of $\Upsilon J/\psi$ production:
\begin{eqnarray*}
d\hat{\sigma}\left(gg\to\Upsilon J/\psi+2\gamma\right) & = & \sum_{J_{b},J_{c}=0,2}\Br\left(\chi_{bJ_{b}}\to\Upsilon\gamma\right)\Br\left(\chi_{cJ_{c}}\to J/\psi\gamma\right)d\hat{\sigma}\left(gg\to\chi_{cJ_{c}}\chi_{cJc}\right),
\end{eqnarray*}
where branching fractions of radiative decays are equal to \cite{Agashe:2014kda}
\begin{eqnarray*}
\Br\left(\chi_{c0}\to J/\psi\gamma\right) & = & 1.3\%,\qquad\Br\left(\chi_{b0}\to\Upsilon(1S)\gamma\right)=1.8\%,\\
\Br\left(\chi_{c1}\to J/\psi\gamma\right) & = & 34\%,\qquad\Br\left(\chi_{b1}\to\Upsilon(1S)\gamma\right)=34\%,\\
\Br\left(\chi_{c2}\to J/\psi\gamma\right) & = & 19\%,\qquad\Br\left(\chi_{b2}\to\Upsilon(1S)\gamma\right)=19\%.
\end{eqnarray*}
Since branching fractions of scalar mesons' radiative decays are small, main contributions come from axial and tensor particles. At partonic energies $\hat{s} = \hat{s}_0$ presented above values give
\begin{eqnarray*}
\hat{\sigma}\left(gg\to\chi_{b}\chi_{c}\to\Upsilon J/\psi+2\gamma\right) & = & 3.4\ \fb.
\end{eqnarray*}
Dependence of $gg\to\chi_{b}\chi_{c}\to\Upsilon J/\psi+2\gamma$ reaction on partonic energy and $\hat{t}$ dependence at $\hat s=\hat{s}_0$ are shown in Fig.\ref{fig:hplotNLO}  by solid line.

\section{Higher Order Corrections\label{sec:NLO}}

\begin{figure}
\includegraphics{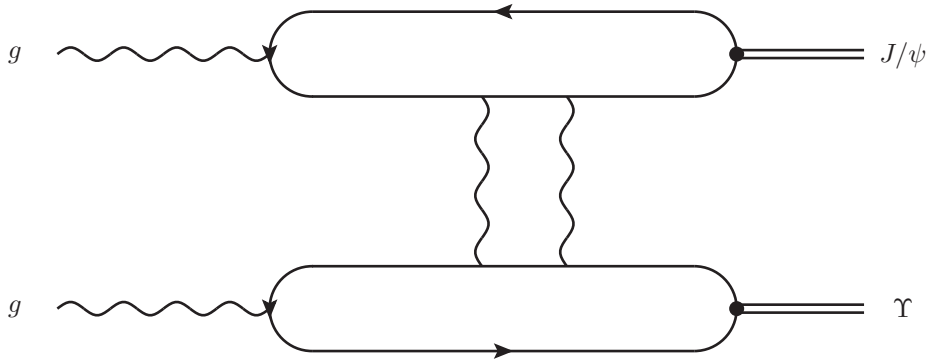}
\caption{
Typical Feynman diagram for $gg\to\Upsilon J/\psi$ reaction
\label{diagNLO}}
\end{figure}

In the Introduction it was mentioned already that one can bypass the prohibition of two vector quarkonia production in gluon-gluon interaction if higher order corrections are considered. Typical Feynman diagram describing the process is shown in Fig.\ref{diagNLO} and in the case of $gg\to2J/\psi$ reaction in was first studied in paper \cite{Kiselev:1988mc} (see also \cite{Ginzburg:1985tp,Ginzburg:1988zy}). According to these articles the cross section of the reaction can be written in the form
\begin{eqnarray}
\frac{d\hat{\sigma}(gg\to\Upsilon J/\psi)}{d\hat{t}} & = & \frac{25\pi^{3}\alpha_{s}^{6}f_{\psi}^{2}f_{\Upsilon}^{2}}{46656m_{b}^{4}m_{c}^{4}}\left[\frac{F(\hat{t})+F(\hat{u})}{2}\right]^{2},\label{eq:dsdtNLO}
\end{eqnarray}
where
\begin{eqnarray}
F(t) & = & \frac{m_{b}m_{c}}{\pi}\int\frac{d^{2}k}{\mathbf{k}^{2}(\mathbf{k}-\mathbf{q})^{2}}\left\{ \frac{1}{1+\tau_{c}}-\frac{1}{1+\tau_{c}\mathbf{r}^{2}}\right\} \left\{ \frac{1}{1+\tau_{b}}-\frac{1}{1+\tau_{b}\mathbf{r}^{2}}\right\} \label{eq:F}
\end{eqnarray}
and notations
\begin{eqnarray*}
\tau_{b,c} & = & \frac{\mathbf{q}^{2}}{4m_{b,c}^{2}},\qquad\mathbf{r}=\frac{2\mathbf{k}-\mathbf{q}}{|\mathbf{q}|}.
\end{eqnarray*}
are introduced. One can easily see that after substitution $m_b=m_c$, $e_b=e_c$ these results agree with those presented in ref.\cite{Ginzburg:1988zy}. The integral (\ref{eq:F}) was calculated numerically, and we found that in the domain $0<|\hat{t}|<10^{4}\ \GeV^{2}$ it can with pretty good accuracy approximated by the relation
\begin{eqnarray*}
F(\hat{t}) & \approx & \frac{Ae^{\alpha\hat{t}}}{1-\hat{t}/M_{0}^{2}},
\end{eqnarray*}
where
\begin{eqnarray*}
A & = & 0.8,\qquad\alpha=2.9\times10^{-3}\ \GeV^{-2},\qquad M_{0}^{2}=3.6\ \GeV^{2}.
\end{eqnarray*}
The mesonic constants in (\ref{eq:dsdtNLO}) are equal to
\begin{eqnarray*}
f_{\psi} & = & 0.4\ \GeV,\qquad f_{\Upsilon}=0.66\ \GeV.
\end{eqnarray*}
Using these parameters we have obtained the number
\begin{eqnarray*}
\hat{\sigma}_{\SPS}^{NLO}\left(gg\to\Upsilon J/\psi\right) & = & 21\ \fb.
\end{eqnarray*}
for integrated partonic cross section at $\hat{s}=\hat{s}_{0}$.
Dependence of this cross section on $\sqrt{\hat s}$ and its $\hat t$ distribution at $\hat s=\hat{s}_0$ are shown in Fig.\ref{fig:hplotNLO} by dotted line. It should be noted however, that these results were obtained in the limit $\hat{s}\gg|\hat{t}|\gg M_{\psi,\Upsilon}^{2},$, so they can be treated only as rough estimates. In our future works we are going to remove this limitation and obtain predictions for shown in Fig.\ref{diagNLO} cross sections in full kinematical region.

The prohibition for $\Upsilon J/\psi$ production in gluonic interaction can be bypassed also if one consider production of colour octet (CO) states \cite{Ko:2010xy}. According to presented in this article results, main contributions in this approximation comes from $gg\to b\overline{b}(^{3}S_{1}^{[1]})c\overline{c}(^{3}S_{1}^{[8]})$, $gg\to b\overline{b}(^{3}S_{1}^{[8]})c\overline{c}(^{3}S_{1}^{[1]})$, and  $gg\to b\overline{b}(^{3}S_{1}^{[8]})c\overline{c}(^{3}S_{1}^{[8]})$ subprocesses. Using presented in \cite{Ko:2010xy} analytical expressions for the cross sections and numerical values
\begin{eqnarray*}
\left\langle O_{1}^{J/\psi}\right\rangle  & = & 1.3\ \GeV^{3},\qquad\qquad\left\langle O_{1}^{\Upsilon}\right\rangle =9.2\ \GeV^{3},\\
\left\langle O_{8}^{J/\psi}\right\rangle  & = & 3.9\times10^{-3}\ \GeV^{3},\qquad\left\langle O_{8}^{\Upsilon}\right\rangle =0.15\ \GeV^{3},
\end{eqnarray*}
for NRQCD matrix elements \cite{Bodwin:2007fz,Kang:2007uv,Braaten:1999qk,Kramer:2001hh}, it is easy to obtain distribution shown by dashed line in Fig.\ref{fig:hplotNLO}. According to this figure $\sqrt{\hat s}$ dependence of different cross sections is quite different, with CO channel giving dominant contribution in region of small invariant mass of the final pair.

\section{Hadronic Cross Sections\label{sec:hadr}}

Let us now consider inclusive $\Upsilon J/\psi$ pair production in proton-proton interaction at LHC energies. The cross section of this process can be easily obtained from presented in previous sections expressions:
\begin{eqnarray}
d\sigma_{\SPS}\left(pp\to\Q_{b}\Q_{c}+X\right) & = & \int\limits _{0}^{1}dx_{1}dx_{2}f_{g}\left(x_{1};\mu^{2}\right)f_{g}\left(x_{2};\mu^{2}\right)d\hat{\sigma}\left(gg\to\Q_{c}\Q_{b}\right),\label{eq:Sigma}
\end{eqnarray}
where $x_{1,2}$ are gluons' momentum fractions, and  $f_{g}(x_{1,2};\mu^{2})$ are distribution functions of these gluons in initial hadrons (parametrisation CTEQ6L \cite{Pumplin:2005rh} will be used in the following), calculated at the scale $\mu^{2}=(M_{\chi_{c}}^{2}+M_{\chi_{b}}^{2})/2$. In LHCb detector conditions, when restrictions $2<y_{cc,bb}<4.5$ are set on the rapidities of final quarkonia, used in our paper model gives the following predictions for cross sections of $\Upsilon J/\psi$ pair production at $\sqrt{s} = 8$ TeV in different channels:
\begin{eqnarray*}
\sigma_{\SPS}\left(pp\to\chi_{b}\chi_{c}+X\to\Upsilon J/\psi+X\right) & = & 0.2\ \pb,\\
\sigma_{\SPS}\left(pp\to\Upsilon J/\psi+X,\NLO\right) & = & 1.5\ \pb,\\
\sigma_{\SPS}\left(pp\to\Upsilon J/\psi,\CO\right) & = & 11.1\ \pb.
\end{eqnarray*}

As it was already mentioned in the Introduction, at LHC energies double parton scattering (DPS) could give noticeable contribution to cross section of the considered process. In this mechanism two final quarkonia are produced in two separate partonic reactions, and in the simplest approximation it can be assumed, that these reactions do not depend on each other. Under these assumptions the cross section of $\Upsilon J/\psi$ pair production is expressed through the cross sections of separate heavy quarkonia production in surprisingly simple relation:
\begin{eqnarray*}
\sigma_{\DPS}\left(\Upsilon J/\psi\right) & = & \frac{\sigma_{\SPS}(\Upsilon)\sigma_{\SPS}(J/\psi)}{\sigma_{eff}}.
\end{eqnarray*}
Using experimental results  $\sigma_{\SPS}(\Upsilon)=0.14\ \mub$, $\sigma_{\SPS}(J/\psi)=1.28\ \mub$ for SPS cross sections and \cite{Aaij:2013yaa} and the value $\sigma_{eff}=14\ \mb$ \cite{Snigirev:2003cq,Korotkikh:2004bz,Novoselov:2011ff}, one can obtain the result
\begin{eqnarray*}
\sigma_{\DPS}\left(pp\to\Upsilon J/\psi+X\right) & = & 12.5\ \pb.
\end{eqnarray*}

\begin{figure}
\includegraphics[width=0.9\textwidth]{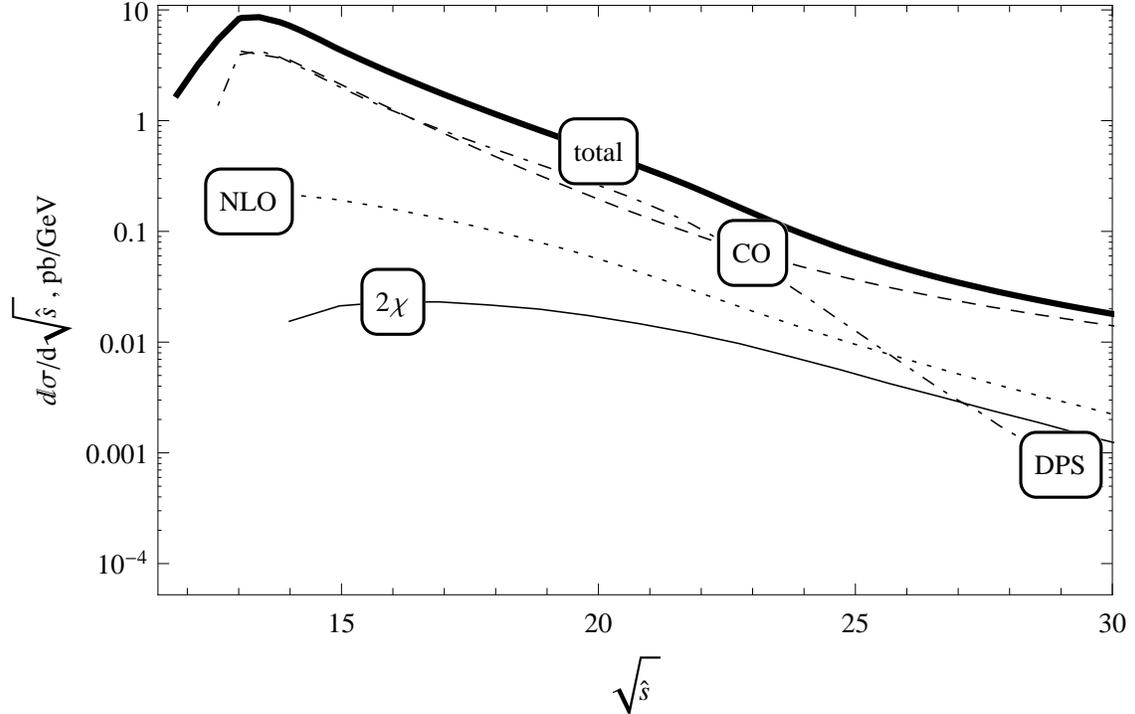}
\caption{
Distribution of $pp\to\Upsilon J/\psi+X$ reaction cross section at LHCb over $\Upsilon J/\psi$-pair invariant mass. Thin solid, dashed, dotted, and dash-dotted  lines correspond to contributions of $\chi_b\chi_c$, CO, NLO, and DPS channels respectively, while thick solid line stands for total cross section\label{fig:mass}}
\end{figure}

\begin{figure}
\includegraphics[width=0.9\textwidth]{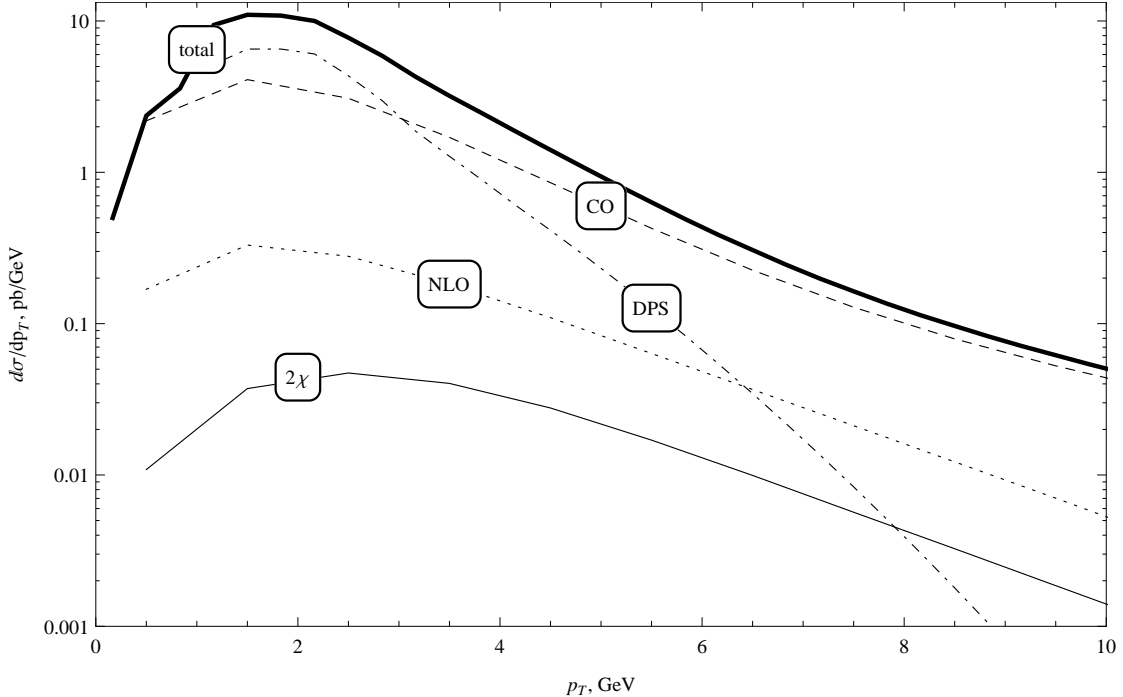}
\caption{
Transverse momentum distribution of $pp\to\Upsilon J/\psi+X$ reaction cross section at LHCb. Notations are same as in Fig.\ref{fig:mass}
\label{fig:pT}}
\end{figure}

It is clear, that this simple mechanism gives main contribution to inclusive $\Upsilon J/\psi$ production at LHCb. Analysis of different distributions, however, could help us to separate contributions of other channels. In Figs.\ref{fig:mass},
\ref{fig:pT}, theoretical predictions for distributions over invariant mass of $\Upsilon J/\psi$ pair and transverse momentum of final quarkonia are shown. On both figures thin solid, dashed, dotted, and dash-dotted  lines correspond to contributions of $\chi_b\chi_c$, CO, NLO, and DPS channels respectively, while thick solid line stands for total cross section. For estimation of DPS distributions we have used Pythia8 generator \cite{Sjostrand:2007gs} with default settings to simulate inpedendent production of $\Upsilon$ and $J/\psi$ mesons and simply combined the resulting events.

\section{Conclusion}

Presented article is devoted to theoretical analysis of inclusive production of heavy quarkonia pair $\Upsilon J/\psi$ at LHCb. It is clear, that at leading order of perturbation theory in colour singlet approximation this process is forbidden, so other mechanisms should be studied. In our paper we consider such channels as double parton scattering, next to leading order corrections, colour octet model and production of vector quarkonia states via radiative decays of $\chi_{b,c}$ mesons. For all these processes theoretical predictions of total cross sections under LHCb detector conditions and distributions over kinematical variables are presented.

According to obtained in our article results, double parton scattering gives main contribution to the cross section of the considered process. From analysis of different distributions. however, it is also possible to separate contributions of other channel, so one can, for example, determine the value of colour-octet NRQCD matrix elements. So, we think that additional theoretical and experimental analysis of inclusive $\Upsilon J/\psi$ pair production at LHC is very interesting and actual task.

The authors would like to thank Dr. Belyaev for fruitful discussions. The work was financially supported by RFBR (grant \#14-02-00096 A) and grant of SAEC "Rosatom" and Helmholtz Association.


\begin{thebibliography}{35}
\expandafter\ifx\csname natexlab\endcsname\relax\def\natexlab#1{#1}\fi
\expandafter\ifx\csname bibnamefont\endcsname\relax
  \def\bibnamefont#1{#1}\fi
\expandafter\ifx\csname bibfnamefont\endcsname\relax
  \def\bibfnamefont#1{#1}\fi
\expandafter\ifx\csname citenamefont\endcsname\relax
  \def\citenamefont#1{#1}\fi
\expandafter\ifx\csname url\endcsname\relax
  \def\url#1{\texttt{#1}}\fi
\expandafter\ifx\csname urlprefix\endcsname\relax\def\urlprefix{URL }\fi
\providecommand{\bibinfo}[2]{#2}
\providecommand{\eprint}[2][]{\url{#2}}

\bibitem[{\citenamefont{Kartvelishvili and
  Esakiya}(1983)}]{Kartvelishvili:1984ur}
\bibinfo{author}{\bibfnamefont{V.}~\bibnamefont{Kartvelishvili}}
  \bibnamefont{and} \bibinfo{author}{\bibfnamefont{S.}~\bibnamefont{Esakiya}},
  \bibinfo{journal}{Yad.Fiz.} \textbf{\bibinfo{volume}{38}},
  \bibinfo{pages}{722} (\bibinfo{year}{1983}).

\bibitem[{\citenamefont{Humpert and Mery}(1983)}]{Humpert:1983yj}
\bibinfo{author}{\bibfnamefont{B.}~\bibnamefont{Humpert}} \bibnamefont{and}
  \bibinfo{author}{\bibfnamefont{P.}~\bibnamefont{Mery}},
  \bibinfo{journal}{Z.Phys.} \textbf{\bibinfo{volume}{C20}},
  \bibinfo{pages}{83} (\bibinfo{year}{1983}).

\bibitem[{\citenamefont{Lansberg and Shao}(2013)}]{Lansberg:2013qka}
\bibinfo{author}{\bibfnamefont{J.-P.} \bibnamefont{Lansberg}} \bibnamefont{and}
  \bibinfo{author}{\bibfnamefont{H.-S.} \bibnamefont{Shao}},
  \bibinfo{journal}{Phys.Rev.Lett.} \textbf{\bibinfo{volume}{111}},
  \bibinfo{pages}{122001} (\bibinfo{year}{2013}), \eprint{arXiv:1308.0474}.

\bibitem[{\citenamefont{Aaij et~al.}(2014)}]{Aaij:2014rms}
\bibinfo{author}{\bibfnamefont{R.}~\bibnamefont{Aaij}} \bibnamefont{et~al.}
  (\bibinfo{collaboration}{LHCb Collaboration}), \bibinfo{journal}{J.Phys.}
  \textbf{\bibinfo{volume}{G41}}, \bibinfo{pages}{115002}
  (\bibinfo{year}{2014}), \eprint{arXiv:1407.5973}.

\bibitem[{\citenamefont{Khachatryan et~al.}(2014)}]{Khachatryan:2014iia}
\bibinfo{author}{\bibfnamefont{V.}~\bibnamefont{Khachatryan}}
  \bibnamefont{et~al.} (\bibinfo{collaboration}{CMS Collaboration}),
  \bibinfo{journal}{JHEP} \textbf{\bibinfo{volume}{1409}}, \bibinfo{pages}{094}
  (\bibinfo{year}{2014}), \eprint{arXiv:1406.0484}.

\bibitem[{\citenamefont{Collaboration}(2013)}]{CMS:2013pph}
\bibinfo{author}{\bibfnamefont{C.}~\bibnamefont{Collaboration}}
  (\bibinfo{collaboration}{CMS Collaboration}) (\bibinfo{year}{2013}).

\bibitem[{\citenamefont{Aaij et~al.}(2012)}]{Aaij:2011yc}
\bibinfo{author}{\bibfnamefont{R.}~\bibnamefont{Aaij}} \bibnamefont{et~al.}
  (\bibinfo{collaboration}{LHCb Collaboration}), \bibinfo{journal}{Phys.Lett.}
  \textbf{\bibinfo{volume}{B707}}, \bibinfo{pages}{52} (\bibinfo{year}{2012}),
  \eprint{arXiv:1109.0963}.

\bibitem[{\citenamefont{Berezhnoy et~al.}(2012)\citenamefont{Berezhnoy,
  Likhoded, Luchinsky, and Novoselov}}]{Berezhnoy:2012xq}
\bibinfo{author}{\bibfnamefont{A.}~\bibnamefont{Berezhnoy}},
  \bibinfo{author}{\bibfnamefont{A.}~\bibnamefont{Likhoded}},
  \bibinfo{author}{\bibfnamefont{A.}~\bibnamefont{Luchinsky}},
  \bibnamefont{and}
  \bibinfo{author}{\bibfnamefont{A.}~\bibnamefont{Novoselov}},
  \bibinfo{journal}{Phys.Rev.} \textbf{\bibinfo{volume}{D86}},
  \bibinfo{pages}{034017} (\bibinfo{year}{2012}), \eprint{arXiv:1204.1058}.

\bibitem[{\citenamefont{Berezhnoy et~al.}(2011)\citenamefont{Berezhnoy,
  Likhoded, Luchinsky, and Novoselov}}]{Berezhnoy:2011xy}
\bibinfo{author}{\bibfnamefont{A.}~\bibnamefont{Berezhnoy}},
  \bibinfo{author}{\bibfnamefont{A.}~\bibnamefont{Likhoded}},
  \bibinfo{author}{\bibfnamefont{A.}~\bibnamefont{Luchinsky}},
  \bibnamefont{and}
  \bibinfo{author}{\bibfnamefont{A.}~\bibnamefont{Novoselov}},
  \bibinfo{journal}{Phys.Rev.} \textbf{\bibinfo{volume}{D84}},
  \bibinfo{pages}{094023} (\bibinfo{year}{2011}), \eprint{arXiv:1101.5881}.

\bibitem[{\citenamefont{Baranov et~al.}(2011)\citenamefont{Baranov, Snigirev,
  and Zotov}}]{Baranov:2011ch}
\bibinfo{author}{\bibfnamefont{S.}~\bibnamefont{Baranov}},
  \bibinfo{author}{\bibfnamefont{A.}~\bibnamefont{Snigirev}}, \bibnamefont{and}
  \bibinfo{author}{\bibfnamefont{N.}~\bibnamefont{Zotov}},
  \bibinfo{journal}{Phys.Lett.} \textbf{\bibinfo{volume}{B705}},
  \bibinfo{pages}{116} (\bibinfo{year}{2011}), \eprint{arXiv:1105.6276}.

\bibitem[{\citenamefont{Novoselov}(2011)}]{Novoselov:2011ff}
\bibinfo{author}{\bibfnamefont{A.}~\bibnamefont{Novoselov}}
  (\bibinfo{year}{2011}), \eprint{arXiv:1106.2184}.

\bibitem[{\citenamefont{Abe et~al.}(1997)}]{Abe:1997xk}
\bibinfo{author}{\bibfnamefont{F.}~\bibnamefont{Abe}} \bibnamefont{et~al.}
  (\bibinfo{collaboration}{CDF Collaboration}), \bibinfo{journal}{Phys.Rev.}
  \textbf{\bibinfo{volume}{D56}}, \bibinfo{pages}{3811} (\bibinfo{year}{1997}).

\bibitem[{\citenamefont{Abazov et~al.}(2010)}]{Abazov:2009gc}
\bibinfo{author}{\bibfnamefont{V.}~\bibnamefont{Abazov}} \bibnamefont{et~al.}
  (\bibinfo{collaboration}{D0 Collaboration}), \bibinfo{journal}{Phys.Rev.}
  \textbf{\bibinfo{volume}{D81}}, \bibinfo{pages}{052012}
  (\bibinfo{year}{2010}), \eprint{arXiv:0912.5104}.

\bibitem[{\citenamefont{Snigirev}(2003)}]{Snigirev:2003cq}
\bibinfo{author}{\bibfnamefont{A.}~\bibnamefont{Snigirev}},
  \bibinfo{journal}{Phys.Rev.} \textbf{\bibinfo{volume}{D68}},
  \bibinfo{pages}{114012} (\bibinfo{year}{2003}), \eprint{hep-ph/0304172}.

\bibitem[{\citenamefont{Berezhnoy et~al.}(2013)\citenamefont{Berezhnoy,
  Likhoded, and Novoselov}}]{Berezhnoy:2012tu}
\bibinfo{author}{\bibfnamefont{A.}~\bibnamefont{Berezhnoy}},
  \bibinfo{author}{\bibfnamefont{A.}~\bibnamefont{Likhoded}}, \bibnamefont{and}
  \bibinfo{author}{\bibfnamefont{A.}~\bibnamefont{Novoselov}},
  \bibinfo{journal}{Phys.Rev.} \textbf{\bibinfo{volume}{D87}},
  \bibinfo{pages}{054023} (\bibinfo{year}{2013}), \eprint{arXiv:1210.5754}.

\bibitem[{\citenamefont{Bodwin et~al.}(1995)\citenamefont{Bodwin, Braaten, and
  Lepage}}]{Bodwin:1994jh}
\bibinfo{author}{\bibfnamefont{G.~T.} \bibnamefont{Bodwin}},
  \bibinfo{author}{\bibfnamefont{E.}~\bibnamefont{Braaten}}, \bibnamefont{and}
  \bibinfo{author}{\bibfnamefont{G.~P.} \bibnamefont{Lepage}},
  \bibinfo{journal}{Phys.Rev.} \textbf{\bibinfo{volume}{D51}},
  \bibinfo{pages}{1125} (\bibinfo{year}{1995}), \eprint{hep-ph/9407339}.

\bibitem[{\citenamefont{Likhoded et~al.}(2012)\citenamefont{Likhoded,
  Luchinsky, and Poslavsky}}]{Likhoded:2012hw}
\bibinfo{author}{\bibfnamefont{A.}~\bibnamefont{Likhoded}},
  \bibinfo{author}{\bibfnamefont{A.}~\bibnamefont{Luchinsky}},
  \bibnamefont{and}
  \bibinfo{author}{\bibfnamefont{S.}~\bibnamefont{Poslavsky}},
  \bibinfo{journal}{Phys.Rev.} \textbf{\bibinfo{volume}{D86}},
  \bibinfo{pages}{074027} (\bibinfo{year}{2012}), \eprint{arXiv:1203.4893}.

\bibitem[{\citenamefont{Likhoded et~al.}(2013)\citenamefont{Likhoded,
  Luchinsky, and Poslavsky}}]{Likhoded:2013aya}
\bibinfo{author}{\bibfnamefont{A.}~\bibnamefont{Likhoded}},
  \bibinfo{author}{\bibfnamefont{A.}~\bibnamefont{Luchinsky}},
  \bibnamefont{and} \bibinfo{author}{\bibfnamefont{S.}~\bibnamefont{Poslavsky}}
  (\bibinfo{year}{2013}), \eprint{arXiv:1305.2389}.

\bibitem[{\citenamefont{Likhoded
  et~al.}(2014{\natexlab{a}})\citenamefont{Likhoded, Luchinsky, and
  Poslavsky}}]{Likhoded:2014gpa}
\bibinfo{author}{\bibfnamefont{A.}~\bibnamefont{Likhoded}},
  \bibinfo{author}{\bibfnamefont{A.}~\bibnamefont{Luchinsky}},
  \bibnamefont{and}
  \bibinfo{author}{\bibfnamefont{S.}~\bibnamefont{Poslavsky}},
  \bibinfo{journal}{Phys.Atom.Nucl.} \textbf{\bibinfo{volume}{77}},
  \bibinfo{pages}{917} (\bibinfo{year}{2014}{\natexlab{a}}).

\bibitem[{\citenamefont{Likhoded
  et~al.}(2014{\natexlab{b}})\citenamefont{Likhoded, Luchinsky, and
  Poslavsky}}]{Likhoded:2014kfa}
\bibinfo{author}{\bibfnamefont{A.}~\bibnamefont{Likhoded}},
  \bibinfo{author}{\bibfnamefont{A.}~\bibnamefont{Luchinsky}},
  \bibnamefont{and}
  \bibinfo{author}{\bibfnamefont{S.}~\bibnamefont{Poslavsky}},
  \bibinfo{journal}{Phys.Rev.} \textbf{\bibinfo{volume}{D90}},
  \bibinfo{pages}{074021} (\bibinfo{year}{2014}{\natexlab{b}}),
  \eprint{arXiv:1409.0693}.

\bibitem[{\citenamefont{Bolotin and Poslavsky}(2013)}]{redberry}
\bibinfo{author}{\bibfnamefont{D.~A.} \bibnamefont{Bolotin}} \bibnamefont{and}
  \bibinfo{author}{\bibfnamefont{S.~V.} \bibnamefont{Poslavsky}},
  \bibinfo{journal}{CoRR} \textbf{\bibinfo{volume}{abs/1302.1219}}
  (\bibinfo{year}{2013}), \urlprefix\url{http://arxiv.org/abs/1302.1219}.

\bibitem[{\citenamefont{Olsson et~al.}(1984)\citenamefont{Olsson, Martin, and
  Peacock}}]{Olsson:1984qe}
\bibinfo{author}{\bibfnamefont{M.}~\bibnamefont{Olsson}},
  \bibinfo{author}{\bibfnamefont{A.~D.} \bibnamefont{Martin}},
  \bibnamefont{and} \bibinfo{author}{\bibfnamefont{A.}~\bibnamefont{Peacock}}
  (\bibinfo{year}{1984}).

\bibitem[{\citenamefont{Kiselev et~al.}(1989)\citenamefont{Kiselev, Likhoded,
  Slabospitsky, and Tkabladze}}]{Kiselev:1988mc}
\bibinfo{author}{\bibfnamefont{V.}~\bibnamefont{Kiselev}},
  \bibinfo{author}{\bibfnamefont{A.}~\bibnamefont{Likhoded}},
  \bibinfo{author}{\bibfnamefont{S.}~\bibnamefont{Slabospitsky}},
  \bibnamefont{and}
  \bibinfo{author}{\bibfnamefont{A.}~\bibnamefont{Tkabladze}},
  \bibinfo{journal}{Yad.Fiz.} \textbf{\bibinfo{volume}{49}},
  \bibinfo{pages}{1681} (\bibinfo{year}{1989}).

\bibitem[{\citenamefont{Olive et~al.}(2014)}]{Agashe:2014kda}
\bibinfo{author}{\bibfnamefont{K.}~\bibnamefont{Olive}} \bibnamefont{et~al.}
  (\bibinfo{collaboration}{Particle Data Group}), \bibinfo{journal}{Chin.Phys.}
  \textbf{\bibinfo{volume}{C38}}, \bibinfo{pages}{090001}
  (\bibinfo{year}{2014}).

\bibitem[{\citenamefont{Ginzburg et~al.}(1987)\citenamefont{Ginzburg, Panfil,
  and Serbo}}]{Ginzburg:1985tp}
\bibinfo{author}{\bibfnamefont{I.}~\bibnamefont{Ginzburg}},
  \bibinfo{author}{\bibfnamefont{S.}~\bibnamefont{Panfil}}, \bibnamefont{and}
  \bibinfo{author}{\bibfnamefont{V.}~\bibnamefont{Serbo}},
  \bibinfo{journal}{Nucl.Phys.} \textbf{\bibinfo{volume}{B284}},
  \bibinfo{pages}{685} (\bibinfo{year}{1987}).

\bibitem[{\citenamefont{Ginzburg et~al.}(1988)\citenamefont{Ginzburg, Panfil,
  and Serbo}}]{Ginzburg:1988zy}
\bibinfo{author}{\bibfnamefont{I.}~\bibnamefont{Ginzburg}},
  \bibinfo{author}{\bibfnamefont{S.}~\bibnamefont{Panfil}}, \bibnamefont{and}
  \bibinfo{author}{\bibfnamefont{V.}~\bibnamefont{Serbo}},
  \bibinfo{journal}{Nucl.Phys.} \textbf{\bibinfo{volume}{B296}},
  \bibinfo{pages}{569} (\bibinfo{year}{1988}).

\bibitem[{\citenamefont{Ko et~al.}(2011)\citenamefont{Ko, Yu, and
  Lee}}]{Ko:2010xy}
\bibinfo{author}{\bibfnamefont{P.}~\bibnamefont{Ko}},
  \bibinfo{author}{\bibfnamefont{C.}~\bibnamefont{Yu}}, \bibnamefont{and}
  \bibinfo{author}{\bibfnamefont{J.}~\bibnamefont{Lee}},
  \bibinfo{journal}{JHEP} \textbf{\bibinfo{volume}{1101}}, \bibinfo{pages}{070}
  (\bibinfo{year}{2011}), \eprint{arXiv:1007.3095}.

\bibitem[{\citenamefont{Bodwin et~al.}(2008)\citenamefont{Bodwin, Chung, Kang,
  Lee, and Yu}}]{Bodwin:2007fz}
\bibinfo{author}{\bibfnamefont{G.~T.} \bibnamefont{Bodwin}},
  \bibinfo{author}{\bibfnamefont{H.~S.} \bibnamefont{Chung}},
  \bibinfo{author}{\bibfnamefont{D.}~\bibnamefont{Kang}},
  \bibinfo{author}{\bibfnamefont{J.}~\bibnamefont{Lee}}, \bibnamefont{and}
  \bibinfo{author}{\bibfnamefont{C.}~\bibnamefont{Yu}},
  \bibinfo{journal}{Phys.Rev.} \textbf{\bibinfo{volume}{D77}},
  \bibinfo{pages}{094017} (\bibinfo{year}{2008}), \eprint{arXiv:0710.0994}.

\bibitem[{\citenamefont{Kang et~al.}(2007)\citenamefont{Kang, Kim, Lee, and
  Yu}}]{Kang:2007uv}
\bibinfo{author}{\bibfnamefont{D.}~\bibnamefont{Kang}},
  \bibinfo{author}{\bibfnamefont{T.}~\bibnamefont{Kim}},
  \bibinfo{author}{\bibfnamefont{J.}~\bibnamefont{Lee}}, \bibnamefont{and}
  \bibinfo{author}{\bibfnamefont{C.}~\bibnamefont{Yu}},
  \bibinfo{journal}{Phys.Rev.} \textbf{\bibinfo{volume}{D76}},
  \bibinfo{pages}{114018} (\bibinfo{year}{2007}), \eprint{arXiv:0707.4056}.

\bibitem[{\citenamefont{Braaten et~al.}(2000)\citenamefont{Braaten, Kniehl, and
  Lee}}]{Braaten:1999qk}
\bibinfo{author}{\bibfnamefont{E.}~\bibnamefont{Braaten}},
  \bibinfo{author}{\bibfnamefont{B.~A.} \bibnamefont{Kniehl}},
  \bibnamefont{and} \bibinfo{author}{\bibfnamefont{J.}~\bibnamefont{Lee}},
  \bibinfo{journal}{Phys.Rev.} \textbf{\bibinfo{volume}{D62}},
  \bibinfo{pages}{094005} (\bibinfo{year}{2000}), \eprint{hep-ph/9911436}.

\bibitem[{\citenamefont{Kramer}(2001)}]{Kramer:2001hh}
\bibinfo{author}{\bibfnamefont{.}~\bibnamefont{Kramer},
  \bibfnamefont{Michael}}, \bibinfo{journal}{Prog.Part.Nucl.Phys.}
  \textbf{\bibinfo{volume}{47}}, \bibinfo{pages}{141} (\bibinfo{year}{2001}),
  \eprint{hep-ph/0106120}.

\bibitem[{\citenamefont{Pumplin et~al.}(2006)\citenamefont{Pumplin, Belyaev,
  Huston, Stump, and Tung}}]{Pumplin:2005rh}
\bibinfo{author}{\bibfnamefont{J.}~\bibnamefont{Pumplin}},
  \bibinfo{author}{\bibfnamefont{A.}~\bibnamefont{Belyaev}},
  \bibinfo{author}{\bibfnamefont{J.}~\bibnamefont{Huston}},
  \bibinfo{author}{\bibfnamefont{D.}~\bibnamefont{Stump}}, \bibnamefont{and}
  \bibinfo{author}{\bibfnamefont{W.}~\bibnamefont{Tung}},
  \bibinfo{journal}{JHEP} \textbf{\bibinfo{volume}{0602}}, \bibinfo{pages}{032}
  (\bibinfo{year}{2006}), \eprint{hep-ph/0512167}.

\bibitem[{\citenamefont{Aaij et~al.}(2013)}]{Aaij:2013yaa}
\bibinfo{author}{\bibfnamefont{R.}~\bibnamefont{Aaij}} \bibnamefont{et~al.}
  (\bibinfo{collaboration}{LHCb collaboration}), \bibinfo{journal}{JHEP}
  \textbf{\bibinfo{volume}{1306}}, \bibinfo{pages}{064} (\bibinfo{year}{2013}),
  \eprint{arXiv:1304.6977}.

\bibitem[{\citenamefont{Korotkikh and Snigirev}(2004)}]{Korotkikh:2004bz}
\bibinfo{author}{\bibfnamefont{V.}~\bibnamefont{Korotkikh}} \bibnamefont{and}
  \bibinfo{author}{\bibfnamefont{A.}~\bibnamefont{Snigirev}},
  \bibinfo{journal}{Phys.Lett.} \textbf{\bibinfo{volume}{B594}},
  \bibinfo{pages}{171} (\bibinfo{year}{2004}), \eprint{hep-ph/0404155}.

\bibitem[{\citenamefont{Sjostrand et~al.}(2008)\citenamefont{Sjostrand, Mrenna,
  and Skands}}]{Sjostrand:2007gs}
\bibinfo{author}{\bibfnamefont{T.}~\bibnamefont{Sjostrand}},
  \bibinfo{author}{\bibfnamefont{S.}~\bibnamefont{Mrenna}}, \bibnamefont{and}
  \bibinfo{author}{\bibfnamefont{P.~Z.} \bibnamefont{Skands}},
  \bibinfo{journal}{Comput.Phys.Commun.} \textbf{\bibinfo{volume}{178}},
  \bibinfo{pages}{852} (\bibinfo{year}{2008}), \eprint{arXiv:0710.3820}.

\end{thebibliography}

\end{document}